\documentclass[12pt]{iopart}
\usepackage[T1]{fontenc}
\usepackage[latin1]{inputenc}
\usepackage{graphicx,amsfonts}
\usepackage{iopams}

\newcommand{\be}{\begin{equation}}
\newcommand{\ee}{\end{equation}}
\newcommand{\bea}{\begin{eqnarray}}
\newcommand{\eea}{\end{eqnarray}}
\newcommand{\bml}{\numparts}
\newcommand{\eml}{\endnumparts}

\newcommand{\ep}{\epsilon}

\eqnobysec

\begin{document}
\title[Diffeomorphism invariance and diffeomorphism generation in the H-L gravity]{Diffeomorphism invariance and diffeomorphism generation in the Ho\v rava-Lifshitz gravity}
\author{F S Bemfica}
\address{Instituto de Física, Universidade de São Paulo\\
Caixa Postal 66318, 05315-970, São Paulo, SP, Brazil}
\eads{fbemfica@fma.if.usp.br}

\begin{abstract}
This paper is intended to study diffeomorphism invariance and diffeomorphism generation in the modified theory of gravity proposed by Ho\v rava. Firstly, we demonstrate that the theory does not lose diffeomorphism invariance due to the parameter $\lambda$, as it was previously believed. However, we show that the presence of terms containing the Levi-Civita symbol in the original proposal of Ho\v rava makes the theory diffeomorphism dependent. By neglecting such terms, what returns fully diffeomorphism invariance to the action, we obtain the equations of motion. Secondly, in the Hamiltonian formalism, we calculate the transformations generated by some of the constraints of the theory. Then, we prove that all diffeomorphisms of General Relativity are generated, on the energy shell, by the constraints of the Ho\v rava-Lifshitz gravity.
\end{abstract}


\maketitle

\section{Introduction}

The modified theory of gravity proposed by Ho\v rava~\cite{Horava2}, also known as Ho\v rava-Lifshitz gravity, has been the focus of great interest in recent times (A status report on the subject can be found in \cite{Sotiriou}. See also \cite{Padilla1,Padilla2,Padilla3,Germani,Horava1,Bogdanos,Henneaux,Pons,Bellorin}.) In his formulation, terms containing higher order in spatial derivatives are added with the intent of modifying the propagator of the theory, turning it finite in the ultraviolet regime. This is done in such a way that terms containing higher order time derivatives are not considered, preventing the emergence of pathologies such as ghosts~\cite{Stelle}. To this end, Ho\v rava imposes an anisotropy in spacetime labeled by a dimensionful running parameter $b$ through the transformations $t\to b^z t$, $x^a\to b x^a$ ($t\in\Re$, $x^a\in\sigma$, $a=1,2,3$), when defined in a foliation $M\cong \Re\times \sigma$. In an attempt to make this anisotropy explicit, causing a partial lose of diffeomorphism invariance, Ho\v rava also introduced a parameter $\lambda $ in the kinetic term of the action.

In this work we will study diffeomorphism invariance and diffeomorphism generation in the theory of Ho\v rava. Taking into account the fact that  ($t,x^a$) are dummy variables of integration, we can perform the inverse transformation $b^z t\to t$, $b x^a\to x^a$ and write the Ho\v rava-Lifshitz action~\cite{Horava2} as
\be
\label{1}
\fl
S=\frac{1}{\kappa}\int_\Re dt\int_\sigma d^3x N\sqrt{q}\left[K^{ab}K_{ab}-\lambda K^2+ R^{(3)}+f(R^{(3)},R^{(3)}_{ab},\dots)\right]\,.
\ee
Here, $R^{(3)}$ is the scalar curvature of the three-dimensional space $\sigma$, $q_{ab}$ is the 3-metric on $\sigma$, while $K_{ab}=(\dot{q}_{ab}-2D_{(a}N_{b)})/2N=(1/2){\mathcal L}_n q_{ab}$\footnote{${\mathcal L}_n$ is the Lie derivative and generates passive diffeomorphisms in the direction of $n$.
We use the notation ${\mathcal L}_v T^{a_1\cdots a_n}_{b_1\cdots b_m}$ to shorten $({\mathcal L}_v T)^{a_1\cdots a_n}_{b_1\cdots b_m}$, where $T$
is an arbitrary tensor field.} is the extrinsic curvature, also known as the second fundamental form, with trace defined as $K=q^{ab}K_{ab}$. We use $\kappa=16\pi G$ with $c=1$, where $G$ is the Newton constant. $N$ and $N^a$ are the Lapse function and Shift vector of the Arnowitt-Deser-Misner (ADM)~\cite{Thiemann,Wald} formalism, respectively. They define the vector field $\partial_t:= Nn+N^a\partial_a$. The unitary timelike vector field $n$ is normal to $\sigma$ and points to the future, thus $N>0$ and $\sqrt{-g}=N\sqrt{q}$. We shall use the notation $g=\det(g_{\mu\nu})$, while $q=\det(q_{ab})$. The torsion free covariant derivative compatible with the 3-metric $q_{ab}$ is $D_a$.

The essential difference between the action in (\ref{1}) and the action of General Relativity (GR) in the ADM formalism is the presence of the dimensionless parameter $\lambda $ and the function $f(R^{(3)},R^{(3)}_{ab},\dots)$. The just mentioned function must contain terms like $(R^{(3)})^2$, $R^{(3)ab}R^{(3)}_{ab}$, $C^{ab}C_{ab}$~\footnote{Despite the fact that the Cotton tensor depends on the constant Levi-Civita symbol $C^{ab}=\varepsilon^{acd}D_c(R^{(3)b}_d-\delta^b_dR^{(3)})$,
the contraction $C^{ab}C_{ab}$ does not.}, among others. We neglect terms depending on the constant symbol of Levi-Civita such as $\varepsilon^{abc}R_{ad}D_bR^d_c$, which does not transform as a tensor and make the theory completely diffeomorphism dependent. In this article we will not stick in the details of $f$, unless it depends on the scalar of curvature $R^{(3)}$, the Ricci tensor $R^{(3)}_{ab}$, and their covariant derivatives $D_a$. Moreover, the most important property of $f$ is this being a scalar. It is worth saying that $\lambda$ and $f$ must satisfy the conditions $f\to0$ and $\lambda\to1$ in the limit where GR apply.

In the section that follows, we will study the diffeomorphism invariance of (\ref{1}). We shall demonstrate that the presence of the parameter $\lambda$ does not affect the diffeomorphism invariance of the theory (See also \cite{Germani}). The presence of terms containing the constant Levi-Civita symbol completely destroy this invariance, while, in the absence of such terms, the theory is diffeomorphism invariant under Diff(M). In section~\ref{sec3}, we calculate the equations of motion for the action in (\ref{1}). The Hamiltonian formalism of the theory is reviewed in section~\ref{sec4}, where we show that the new Hamiltonian and vector constraints are directly related to the equations of motion of the theory. Then, we dedicate ourselves to calculate the transformations generated by these constraints. We prove that, on the energy shell, they are the generators of the diffeomorphisms of $M$, as in GR. In section~\ref{Conclusions} are the conclusions.

\section{Diffeomorphism invariance}
\label{sec2}

It has been argued~\cite{Horava2} that the modified theory given in (\ref{1}), in the presence of $\lambda$, is restricted to a subgroup Diff$_{\mathcal F}(M)\subset\mathrm{Diff}(M)$ of diffeomorphisms. We shall demonstrate that the theory described by (\ref{1}) is, indeed, invariant throughout all the diffeomorphisms of the GR group Diff$(M)$. In other words, the theory is independent of coordinates. In \cite{Horava2}, the argument imposed is that a theory must have a Lagrangian density $\mathcal{L}$ invariant under active diffeomorphisms (that changes the points but not the differentiable structure of $M$) of the type $p\to p^\prime=p+\delta p\Rightarrow \delta x^\mu(p) = x^\mu(p^\prime)-x^\mu(p):= \epsilon^\mu(t(p),x(p))$ [$x^\mu=(t,x)$.] Here, $p\to p^\prime$ is an active diffeomorphism $M\to M$. We must be aware that the Lagrangian density of any theory may not be invariant under such active diffeomorphisms, that is to say $\Delta_\epsilon {\mathcal L}={\mathcal L}(x^\mu+\epsilon^\mu)-{\mathcal L}(x^\mu)=\epsilon^\mu\partial_\mu{\mathcal L}(x^\mu)\ne0$. This property, in turns, prevents the Lagrangian density ${\mathcal L}$ of being a constant, independent of the spacetime coordinates $x^\mu = (t,x)$ and, as a consequence, of the points $p\in M$. In fact, a diffeomorphism is an isomorphism in the category of smooth manifolds. A bijective map $\varphi$ between differentiable manifolds is a diffeomorphism if $\varphi$ and its inverse are differentiable~\cite{Senhoras}. In the particular case of GR, Diff(M) is the set of maps $\varphi\,:\, M\to M$. Passive diffeomorphisms, or simply diffeomorphisms, includes the changes of parametrization (coordinate systems) of the points of M. Thus, the invariance we are talking about must be present in the transformation $\varphi_*\,:\,{\mathcal L}(x^\mu)\to{\mathcal L}^\prime(x^{\prime\mu})(p)={\mathcal L}(x^\mu)(p)$, keeping the action (\ref{1}) invariant for different parametrizations $x$ and $x^\prime$ of each point $p\in M$. It is important to say that, although the Lagrangian may not be invariant under active diffeomorphism, its action is \cite{Thiemann}.

Let us follow the steps in \cite{Thiemann,Wald} by performing an immersion of $\sigma$ into $M$ through an arbitrary diffeomorphism $X\,:\,\Re\times\sigma\to M;\,\sigma\to \Sigma_t;\,(t,x)\to X(t,x):=X_t(x)$. The spacetime is foliated in a spacelike hypersurface $\Sigma_t$. Any diffeomorphism $\varphi\in \mathrm{Diff}(M)$ has the form $\varphi=X^\prime\circ X^{-1}$, where $X$ and $X^\prime$ are two distinct foliations related by the diffeomorphism $X^\prime=\varphi\circ X$. Thus, since the action in (\ref{1}) is invariant under the immersion $X$, it will be invariant under any diffeomorphism $\varphi\in \mathrm{Diff}(M)$. We define the parametrization
\be
\label{2}
\fl
t^\mu(X):= (\partial_t)^\mu=\left.\frac{\partial X^\mu(t,x)}{\partial t}\right|_{X(t,x)=X}=N(X)n^\mu(X)+N^\mu(X):=X^\mu_t(X)\,,
\ee
where $n^\mu N_\mu=0$ and $n^\mu v_\mu=0$ if $v$ is a covector on $\Sigma_t$. The inverse of (\ref{2}), subtracted from the  relation $dt(\partial_t)=t^\mu\nabla_\mu t=1$, is given by $\nabla_\mu t(X)=-n_\mu(X)/N(X):=(X^{-1})^t_\mu$, which is the covector normal to $\Sigma_t$. Here, $\nabla_\mu$ is the torsion free metric preserving covariant derivative compatible with $g_{\mu\nu}$ instead of $q_{\mu\nu}$. It is convenient to define the following three spacial vector fields of $\Sigma_t$
\be
\label{3}
X^\mu_a:=\left.\frac{\partial X^\mu}{\partial x^a}\right|_{X(t,x)=X}\,.
\ee
Being $X$ a diffeomorphism, (\ref{3}) must have an inverse. Then, we can write
\bea
\label{4}
q^{ab}(t,x)&=& (X^{-1})^a_\mu(X^{-1})^b_\nu q^{\mu\nu}(X(t,x))\,,\nonumber\\
K_{ab}(t,x)&=&X^\mu_aX^\nu_b K_{\mu\nu}(X(t,x))\,.
\eea
This, in turns, enables one to find
\bea
\label{5}
\fl K(x,t)=(q^{ab}K_{ab})(t,x)=(X^{-1})^a_\mu (X^{-1})^b_\nu X^\alpha_aX^\beta_b (q^{\mu\nu}K_{\alpha\beta})(X(t,x))\nonumber\\
=(q^{\mu\nu}K_{\mu\nu})(X(t,x))=K(X(t,x))\,,\nonumber\\
\fl (K^{ab}K_{ab})(t,x)=(X^{-1})^a_\mu (X^{-1})^b_\nu X^\alpha_aX^\beta_b (K_{\alpha\beta}K^{\mu\nu})(X(t,x))\nonumber\\
=(K_{\mu\nu}K^{\mu\nu})(X(t,x))\,.
\eea
The above calculations where performed by taking into account the identity
\be
\label{6}
(X^{-1})^b_\nu X^\beta_b=\delta^\beta_\nu-(X^{-1})^t_\nu X^\beta_t=\delta^\beta_\nu+\frac{n_\nu}{N}t^\beta\,,
\ee
together with $n_\mu q^{\mu\nu}=n_\mu K^{\mu\nu}=0$. In order to save space, from now on we will eventually leave aside the label $X$. The first fundamental form is written as
\be
\label{7}
q_{\mu\nu}=g_{\mu\nu}+n_\mu n_\nu\,.
\ee
One can easily show that $K_{\mu\nu}=(X^{-1})^a_\mu(X^{-1})^b_\nu K_{ab}=q^\alpha_{(\mu} q^\beta_{\nu)}\nabla_\beta n_\alpha={\mathcal L}_nq_{\mu\nu}/2$ [The indexes symmetrization $T_{(a}U_{b)}:=(1/2)(T_aU_b+T_bU_a)$.] Here, the Lie derivative ${\mathcal L}_v$ generates active diffeomorphisms in the direction of the vector field $v$. The covariant derivative compatible with $q_{\mu\nu}$ is defined by $D_\mu f=q^\alpha_\mu \nabla_\mu f$, and $D_\mu v_\nu=q^\alpha_\mu q^\beta_\nu \nabla_\beta v_\alpha$, where $f$ must be a smooth function on $\Sigma_t$ and $n^\mu v_\mu=0$. The same result holds for arbitrary tensor fields defined on this hypersurface. The reader may verify that
\bea
\label{7-1}
D_a v_b&=&X^\mu_a X^\nu_b D_\mu v_\nu=X^\mu_a X^\nu_b q^\alpha_\mu q^\beta_\nu \nabla_\alpha v_\beta\,,\quad v_X\in T_X\Sigma_t\nonumber\\
&=&\partial_a v_b - \Gamma^{(3)c}_{ab}v_c\,.
\eea
This result can be extended to any tensor field in $\Sigma_t$, what enables one to write $R^{(3)}(t,x)=R^{(3)}(X(t,x))$, $R^{(3)}_{ab}=X^\mu_aX^\nu_b R^{(3)}_{\mu\nu}(X(t,x))$, and so on. Details regarding these and other calculations are in \cite{Thiemann,Wald}. We argue that the function $f(R^{(3)},R^{(3)}_{ab},\dots)$, in the absence of terms containing the constant Levi-Civita symbol, is a scalar. Thus,  $f(R^{(3)},R^{(3)}_{ab},\dots)(t,x)=f(R^{(3)},R^{(3)}_{\mu\nu},\dots)(X(t,x))$.

What remains to be done is to write $N\sqrt{q}=\sqrt{-g(t,x)}$ on the pushforward $X_*(dtd^3x\sqrt{-g(t,x)})=d^4X\sqrt{-g(X)}$. Collecting all the above results, the action in (\ref{1}) takes the form
\be
\label{8}
S=\frac{1}{\kappa}\int_M d^4X \sqrt{-g}\left[K^{\mu\nu}K_{\mu\nu}-\lambda K^2+ R^{(3)}+f(R^{(3)},R^{(3)}_{\mu\nu},\dots)\right]\,,
\ee
for any diffeomorphism $X$. This demonstrates that the introduction of $\lambda$ in the theory of Ho\v rava does not affect its diffeomorphism invariance.

At this moment, it is important to emphasize that a term like $\varepsilon^{abc}R_{ad}D_bR^d_c$ is clearly dependent of the diffeomorphism we choose. In other words, although the Ricci tensor transforms as a tensor of rank 2, the Levi-Civita symbol $\varepsilon^{abc}$, which is a constant, does not, and the theory loses all its diffeomorphism invariance, even Diff$_{\mathcal F}(M)$, as argued by Ho\v rava~\cite{Horava2}.

In the next sections we will turn to the problem of diffeomorphism transformations and show that the theory is not only diffeomorphism invariant, as seen here, but also possess its constraints as the generators of the active diffeomorphisms of GR.

\section{The equations of motion}
\label{sec3}

From now on we set, for simplicity,
\be
\kappa=1\,.
\ee
To calculate the equations of motion for the action in (\ref{1}), we first rewrite (\ref{8}) using the reverse procedure of the ADM formalism, namely,
\be
\label{9}
S=\int_M d^4X \sqrt{-g}\left[R^{(4)}+\left(1-\lambda\right) K^2+f(R^{(3)},R^{(3)}_{\mu\nu},\dots)\right]\,.
\ee
The identity
\be
\label{10}
R^{(4)}=R^{(3)}+\left[K^{\mu\nu}K_{\mu\nu}-K^2\right]-2\nabla_\mu\left(n^\nu\nabla_\nu n^\mu-n^\mu\nabla_\nu n^\nu\right)\,,
\ee
from the Gauss-Codazzi equation~\cite{Wald}, was taken into account. The scalar curvature $R^{(4)}$, defined on $M$, stems from the definition for the curvature tensor $[\nabla_\mu\,,\,\nabla_\nu]v_\alpha=R^{(4)\beta}_{\;\;\;\mu\nu\alpha}v_\beta$. The last term in (\ref{10}) is a surface term and has been neglected. By performing the inverse diffeomorphism $X^{-1}\,:\, M\to \Re\times\sigma$ on (\ref{9}) we obtain
\be
\label{11}
\fl S=\int_\Re dt\int_\sigma d^3x \sqrt{-g(t,x)}\left[R^{(4)}+\left(1-\lambda\right) K^2+f(R^{(3)},R^{(3)}_{ab},\dots)(t,x)\right]\,.
\ee

Let us recall some important relations. In the coordinate system $(t,x^a)$, the normal vector field $n$ has components $n^\mu=(1/N,-N^a/N)$, while $n_\mu=(-N,0)$. The metric on $\sigma$ is given by $q_{ab}$ ($q^a_b=\delta^a_b$), while $q^\mu_\nu$ may project any vector field on $M$ into $\sigma$. Again, we write $\sqrt{-g}=N\sqrt{q}$, with $q=\det(q_{ab})$. The remaining components of  $q_{\mu\nu}(q^{\mu\nu})$ are: $q_{ta}=N_a$, $q_{tt}=N^aN_a$ ($q^{tt}=q^{ta}=0$,) being $n^\mu q_{\mu\nu}=0$ and $g_{\mu\nu}=q_{\mu\nu}-n_\mu n_\nu$. The properties of the projectors $q$ and $n\otimes n$ enable us to set down any vector field $v$ on $M$ as
\be
\label{11-2}
v=(v+g(v,n)n)-g(v,n)n:=\hat{v}n+\hat{v}^a\partial_a\,,
\ee
where
\bml
\bea
\label{11-3}
\hat{v}&=&-n_\mu v^\mu=N v^t\,,\label{11-3a}\\
\hat{v}^a&=&q^a_\mu v^\mu=v^a-N^av^t\,.\label{11-3b}
\eea
\eml
Above, $\hat{v}$ is the $v$ component normal to $\sigma$, while its tangent component is $\hat{v}^a$, both written in the nonholonomic basis $e_\mu=(n,\partial_a)$.

We return to the variation of the action in (\ref{11}) and write down
\bea
\label{11-4}
\delta S&:=&\left.\frac{d S[g+s\delta g]}{ds}\right|_{s=0}\nonumber\\
&=&\int_\Re dt\int_\sigma d^3x\left\{\left[R^{(4)}+(1-\lambda)K^2+f(R^{(3)},R^{(3)}_{ab},\dots)\right]\sqrt{q}\delta N\right.\nonumber\\
&+&\frac{1}{2}\left[R^{(4)}+(1-\lambda)K^2+f(R^{(3)},R^{(3)}_{ab},\dots)\right]N\sqrt{q}q^{ab}\delta q_{ab}\nonumber\\
&+&\sqrt{-g}\left.\left[-R^{\mu\nu}\delta g_{\mu\nu}+2(1-\lambda)K\delta K+\delta f(R^{(3)},R^{(3)}_{ab},\dots)\right]\right\}\,.
\eea
The relation above has been calculated by means of the equality $\delta\sqrt{q}=(1/2)\sqrt{q}q^{\mu\nu}\delta q_{\mu\nu}=(1/2)\sqrt{q}q^{ab}\delta q_{ab}$, together with $q^{ta}=q^{tt}=0$. After we take the variation $\delta R^{(4)}=-R^{(4)}\delta g_{\mu\nu}+\nabla^\alpha\left(\delta_\alpha^{(\mu}\nabla^{\nu)}-g^{\mu\nu}\nabla_\alpha\right)\delta g_{\mu\nu}$, we discarded the surface term in parenthesis.

We must express the total variation of $S$ in terms of the independent variables $N$, $N^a$, and $q_{ab}$, separately. Thereunto, it is necessary an analyzes of each term in the last equality in (\ref{11-4}). We may begin writing
\bea
\label{12}
-R^{(4)\mu\nu}\delta g_{\mu\nu}&=&-R^{(4)\alpha\beta}\delta^\mu_\alpha\delta^\nu_\beta\delta g_{\mu\nu}
=-R^{(4)\alpha\beta}(q^\mu_\alpha-n^\mu n_\alpha)(q^\nu_\beta-n^\nu n_\beta)\delta g_{\mu\nu}\nonumber\\
&=&\left(-R^{(4)\alpha\beta}q^a_\alpha q^b_\beta+2R^{(4)\alpha\beta}q^a_\alpha n_\beta n^b\right)\delta q_{ab}
+\frac{2}{N}R^{(4)\alpha\beta}q^a_\alpha n_\beta \delta N_a\nonumber\\
&+&\frac{2}{N}R^{(4)\alpha\beta}n_\alpha n_\beta\delta N\,.
\eea
Now, it is needed to calculate the variation of the trace $K$. This can be easily achieved expressing $K$ as $K=\nabla_\mu n^\mu$~\cite{Thiemann,Wald}, then
\be
\label{13}
\delta K=\nabla_\mu \delta n^\mu+\frac{1}{2}n^\alpha g^{\lambda\beta}\nabla_\alpha\delta g_{\lambda\beta}\,.
\ee
The identity $n_\alpha\delta n^\alpha=-n^\alpha\delta n_\alpha$ helps us to find
\bea
\label{14}
\delta n^\mu&=&q^\mu_\alpha\delta n^\alpha -n^\mu n_\alpha\delta n^\alpha\nonumber\\
&=&-n^\alpha q^{\mu a}\delta q_{a\alpha}-\frac{n^\mu}{N}\delta N\,.
\eea
Equations.~(\ref{13}) and (\ref{14}) are essential to calculate the integral that follows,
\bea
\label{15}
&&\int_\Re dt\int_\sigma d^3x \sqrt{-g}K\delta K\nonumber\\
&&=\int_\Re dt\int_\sigma d^3x \sqrt{-g}\left[-\delta n^\mu\nabla_\mu K
-\frac{1}{2}g^{\lambda\beta}\left(K^2+\nabla_n K\right)\delta g_{\lambda\beta}\right]\nonumber\\
&&=\int_\Re dt\int_\sigma d^3x N\sqrt{q}\Bigg[n^\alpha \delta q_{a\alpha}D^a K+
\frac{1}{N}\delta N \,\nabla_n K\nonumber\\
&&\quad-\frac{1}{2}\left(K^2+\nabla_n K\right)g^{\lambda\beta}\delta g_{\lambda\beta}\Bigg]\nonumber\\
&&=\int_\Re dt\int_\sigma d^3x N\sqrt{q}\left\{\frac{\delta N}{N}(-K^2)+\frac{\delta N_a}{N}(D^aK)\right.\nonumber\\
&&+\left.\delta q_{ab}\left[n^{(a}D^{b)}-\frac{ q^{ab}}{2}\left(K^2+\nabla_n K\right)\right]\right\}\,.
\eea
In the first equality above we have performed an integral by parts, neglecting surface terms. We also used the equality $q^{\mu a}\nabla_\mu K=D^a K$ and, afterwards, applied the relation for $g$ given by
\bea
\label{16}
g^{\lambda\beta}\delta g_{\lambda\beta}&=&q^{\lambda\beta}\delta q_{\lambda\beta}+n^\alpha n^\beta\delta(n_\alpha n_\beta)\nonumber\\
&=&q^{ab}\delta q_{ab}-2n^\alpha\delta n_\alpha=q^{ab}\delta q_{ab}+\frac{2}{N}\delta N\,.
\eea

The remaining term to be analyzed in (\ref{11-4}) involves $f(R^{(3)},R^{(3)}_{ab},\dots)$ and needs a special attention. We do not want nor need to treat the exact form of $f$. So, we begin by studying the particular case $w=w(R^{(3)})$. We know that $R^{(3)}$ depends only on $q_{ab}$ an $D_a$ through $[D_a,D_b]v_c=R^{(3)d}_{\;\;\; abc}v_d$, then, the integral
\bea
\label{17}
&& \int_\sigma d^3x N\sqrt{q}\delta w(R^{(3)})\nonumber\\
&&=\int_\sigma d^3x N\sqrt{q}\frac{\partial w(R^{(3)})}{\partial R^{(3)}}\left(-R^{(3)ab}+D^aD^b-q^{ab}D^cD_c\right)\delta q_{ab}\nonumber\\
&&=\int_\sigma d^3x \sqrt{q}\delta q_{ab}\left(-R^{(3)ab}+D^aD^b-q^{ab}D^cD_c\right)\left(N\frac{\partial w(R^{(3)})}{\partial R^{(3)}}\right)\nonumber\\
&&:=\int_\sigma d^3x \sqrt{q}\delta q_{ab} \left[{\mathcal D}_w (N)\right]^{ab}\,,
\eea
where the effect of the variation of $w(R^{(3)})$ inside the integral was summarize, up to surface terms, by the tensor of rank 2
\be
\label{18}
\left[{\mathcal D}_w (N)\right]^{ab}=\left(-R^{(3)ab}+D^aD^b-q^{ab}D^cD_c\right)\left(N\frac{\partial w(R^{(3)})}{\partial R^{(3)}}\right)\,.
\ee
Observe that $\left[{\mathcal D}_w (N)\right]^{ab}$ transforms as a tensor and $n_\mu(X)\left[{\mathcal D}_w (N(X))\right]^{\mu\nu}=0$ in any  diffeomorphism $X$. This result will be extended to the case of (\ref{11-4}). As we imposed, the function $f(R^{(3)},R^{(3)}_{ab},\dots)$ is a covariant combination of the Ricci tensor, scalar curvature and their covariant derivatives. This enables us to write
\bea
\label{19}
&&\int_\sigma d^3x N\sqrt{q}\delta f(R^{(3)},R^{(3)}_{ab},\dots)\nonumber\\
&&=\int_\sigma d^3x N\sqrt{q}\left(\frac{\partial f(R^{(3)},R^{(3)}_{ab},\dots)}{\partial R^{(3)}}\delta R^{(3)}
+\frac{\partial f(R^{(3)},R^{(3)}_{ab},\dots)}{\partial R^{(3)}_{ab}}\delta R^{(3)}_{ab}\right.\nonumber\\
&&\left.+\frac{\partial f(R^{(3)},R^{(3)}_{ab},\dots)}{\partial D_cR^{(3)}_{ab}}\delta (D_cR^{(3)}_{ab})+\cdots\right)\nonumber\\
&&:=\int_\sigma d^3x \sqrt{q}\delta q_{ab}\left[{\mathcal D}_f (N)\right]^{ab}\,,
\eea
where $\left[{\mathcal D}_f (N)\right]^{ab}$ must be a tensor of rank 2 with the property $n_\mu(X)\left[{\mathcal D}_f (N(X))\right]^{\mu\nu}$ in any diffeomorphism $X$. The above result takes into account that any variation of the Ricci tensor, Ricci scalar, or of the Christoffer symbol $\Gamma^{(3)a}_{bc}$, will generate covariant derivatives of $\delta q_{ab}$, that can be eliminated by performing some integrations by parts and by dropping surface terms.

Collecting the results of (\ref{12}), (\ref{15}), and (\ref{19}) we obtain
\bml
\label{20}
\bea
\delta S&=&\int_\Re dt\int_\sigma d^3x\sqrt{q}\Bigg\{\nonumber\\
&&\left[2R^{(4)\alpha\beta}n_\alpha n_\beta +R^{(4)} -(1-\lambda)K^2+f(R^{(3)},R^{(3)}_{ab},\dots)\right]\delta N\label{20a}\\
&&+\Bigg[-R^{(4)\alpha\beta}q^a_\alpha q^b_\beta+2R^{(4)\alpha\beta}q^a_\alpha n_\beta n^b\nonumber\\
&&\quad+\frac{1}{2}q^{ab}R^{(4)}+\frac{1}{2}\left(-(1-\lambda)K^2+f(R^{(3)},R^{(3)}_{ab},\dots)\right)\nonumber\\
&&\left.\quad+(1-\lambda)\left(2n^{(a}D^{b)}K-q^{ab}\nabla_n K\right)+\frac{1}{N}\left[{\mathcal D}_f(N)\right]^{ab}\right]N\delta q_{ab}\label{20b}\\
&&+2\left[R^{(4)\alpha\beta}q^a_\alpha n_\beta+(1-\lambda)D^aK\right]\delta N_a\Bigg\}\,.\label{20c}
\eea
\eml
The terms in brackets must cancel separately so that $\delta S=0$. We claim that the above result can be compactified as follows
\bea
\label{21}
\fl\tilde{G}^{\alpha\beta}:=G^{\alpha\beta}-\frac{1}{2}g^{\alpha\beta}\left[-(1-\lambda)K^2+f(R^{(3)},R^{(3)}_{\mu\nu},\dots)\right]\nonumber\\
-(1-\lambda)n^{(\alpha}D^{\beta)}K+q^{\alpha\beta}(1-\lambda)\nabla_n K-\frac{1}{N}\left[{\mathcal D}_f(N)\right]^{\alpha\beta}=0\,,
\eea
where $G^{\alpha\beta}=R^{(4)\alpha\beta}-(1/2)g^{\alpha\beta}R^{(4)}$ is the Einstein tensor. The reader may verify that the term in brackets in (\ref{20a}) corresponds to $2n_\alpha n_\beta\tilde{G}^{\alpha\beta}$, while the one in (\ref{20b}) is equals to $(q^a_\alpha q^b_\beta-2q^{(a}_\alpha n^{b)}n_\beta)\tilde{G}^{\alpha\beta}$, and, the  remaining term in (\ref{20c}) can be achieved by performing $2q^a_\alpha n_\beta\tilde{G}^{\alpha\beta}$. Thus, as we had claimed, (\ref{21}) is the covariant equation of motion of the Ho\v rava-Lifshitz theory.

\section{Hamiltonian formalism and diffeomorphism generation}
\label{sec4}

Before looking at the transformations generated by the Hamiltonian and vector constraints, we must write the Hamiltonian formalism of the theory. The procedure is similar to the $\lambda R$ theory (where $f(R^{(3)},R^{(3)}_{ab},\dots)=0$) studied in \cite{Henneaux,Pons,Bellorin}. Without going into the details, we write the total Hamiltonian~\cite{Dirac1,Thiemann}
\bea
\label{IV-1}
H&=&H(N)+H_a(N^a)+C(\gamma)+C_a(\gamma^a)\nonumber\\
&=&\int_\sigma d^3x\left(N{\mathcal H}+N^a{\mathcal H}_a+\gamma{\mathcal C}+\gamma^a{\mathcal C}_a\right)\,,
\eea
where
\bml
\label{IV-2}
\bea
{\mathcal H}&=&\frac{1}{\sqrt{q}}\left[\pi^{ab}\pi_{ab}-\tilde{\lambda}\pi^2\right]-\sqrt{q}\left(R^{(3)}+
f(R^{(3)},R^{(3)}_{ab},\dots)\right)\,,\label{IV-2a}\\
{\mathcal H}_a&=&-2 D_b \pi^b_a\,\label{IV-2b}\\
\pi^{ab}&=&\frac{\delta L}{\delta \dot{q}_{ab}}=\sqrt{q}\left(K^{ab}-\lambda q^{ab}K\right)\,,\label{IV-2c}\\
{\mathcal C}&=&\Pi=\frac{\delta L}{\delta \dot{N}}\approx0\,,\label{IV-2d}\\
{\mathcal C}_a&=&\Pi_a=\frac{\delta L}{\delta \dot{N}^a}\approx0\,.\label{IV-2e}
\eea
\eml
We use the definition $\tilde{\lambda}:=\lambda/(3\lambda-1)$. Following the formalism of Dirac~\cite{Dirac1}, ${\mathcal C}$ and ${\mathcal C}_a$ are primary constraints, and their persistence in time lead us to the secondary constraints ${\mathcal H}\approx0$ and ${\mathcal H}_a\approx0$, also known as the Hamiltonian and vector constraints, respectively. In (\ref{IV-1}), $N$ and $N^a$ depends on the spacetime variables $(t,x)$, being coordinates of the phase space $\Gamma=\{q_{ab},N,N^a,\pi^{ab},\Pi,\Pi_a\}$, while $\gamma$ and $\gamma^a$ are Lagrange multipliers. We will denote by $\bar{\Gamma}$ the reduced phase space, where the constraint equations ${\mathcal H}={\mathcal H}_a={\mathcal C}={\mathcal C}_a=0$ must hold. We also define the equal time Poisson brackets for arbitrary functions $F,\,F^\prime\in\Gamma$
\be
\label{IV-3}
\left\{F,F^\prime\right\}:=\int_\sigma d^3y \left(\frac{\delta F}{\delta Q(y)}\frac{\delta F^\prime}{\delta P(y)}-\frac{\delta F}{\delta P(y)}\frac{\delta F^\prime}{\delta Q(y)}\right)\,.
\ee
The reader may perceive that the short hand notation $Q=(q_{ab},N,N^a)$ for the fields and $P=(\pi^{ab},\Pi,\Pi_a)$ for their canonically conjugate momenta has been applied, together with an implicit index summation. At this point, we can identify the relation of the equations of motion with the Hamiltonian and vector constraints. The identities $n_\mu n_\nu G^{\mu\nu}=-(1/2)(K^{ab}K_{ab}-K^2-R^{(3)})$ and $n^\mu q^a_\nu G^{\mu\nu}=D_b(K^{ab}-q^{ab}K)$~\cite{Thiemann,Wald}, together with (\ref{IV-2c}) and the equations of motion $\tilde{G}^{\mu\nu}$ in (\ref{21}), enable us to find
\bml
\label{IV-4}
\bea
n_\mu n_\mu\tilde{G}^{\mu\nu}&=&-\frac{1}{2\sqrt{q}}{\mathcal H}\,,\label{IV-4a}\\
q^a_\mu n_\nu\tilde{G}^{\mu\nu}&=&-\frac{1}{2\sqrt{q}}{\mathcal H}^a\,.\label{IV-4b}
\eea
\eml
In other words, just as in GR, the Ho\v rava-Lifshitz theory on shell defines the reduced phase space $\bar{\Gamma}$.

We shall be concentrated now on the transformations generated by the Hamiltonian and vector constraints. We must verify that these constraints are the generators of the diffeomorphism from Diff(M).
Let us begin by quoting the active diffeomorphism $\delta x^\mu=\epsilon^\mu(t,x)=(\epsilon^t,\epsilon^a)$. We may write $\ep$ in the nonholonomic basis $e_\mu=(n,\partial_a)$ as $\ep=\hat{\ep}n+\hat{\ep}^a\partial_a$, where $\hat{\ep}=N\ep^t$ and $\hat{\ep}^a=\ep^a-N^a\ep^t$. From the active diffeomorphism on $g_{\mu\nu}$ given by ${\mathcal L}_\ep g_{\mu\nu}=\ep^\alpha\partial_\alpha g_{\mu\nu}+2g_{\alpha(\mu}\partial_{\nu)}\ep^\alpha$, it is easy to show that~\cite{Thiemann,Wald}
\bml
\label{IV-6}
\bea
{\mathcal L}_{\hat{\ep}n}q_{ab}&=&\hat{\ep}{\mathcal L}_nq_{ab}=2\hat{\ep}K_{ab}\,,\label{IV-6a}\\
{\mathcal L}_{\hat{\ep}^c\partial_c}q_{ab}&=&2 D_{(a}\hat{\ep}_{b)}\,.\label{IV-6b}
\eea
\eml
Also, it can be verified that
\bml
\label{IV-7}
\bea
\delta_{\hat{\ep}n}q_{ab}&:=&\left\{q_{ab},H(\hat{\ep})\right\}\nonumber\\
&=&\frac{2\hat{\ep}}{\sqrt{q}}\left(\pi_{ab}-\tilde{\lambda}q_{ab}\pi\right)=2\hat{\ep}K_{ab}\,,\label{IV-7a}\\
\delta_{\hat{\ep}^c\partial_c}q_{ab}&:=&\left\{q_{ab},H_c(\hat{\ep}^c)\right\}\nonumber\\
&=&2 D_{(a}\hat{\ep}_{b)}\,.
\eea
\eml
The above equations reflect the fact that the Hamiltonian and vector constraints are the generators of the diffeomorphisms given in (\ref{IV-6a}) and (\ref{IV-6b}). It now remains to verify the active diffeomorphisms for the canonically conjugate momenta $\pi^{ab}$, since the transformations that generate the active diffeomorphisms on $N$ and $N^a$ are performed by the smeared functions $C({\mathcal L}_\ep N)$ and $C_a({\mathcal L}_\ep N^a)$, respectively. By considering the active diffeomorphism
\be
\label{IV-8}
{\mathcal L}_{\hat{\ep}^c\partial_c}K^{ab}=\hat{\ep}^cD_c K^{ab}-2K^{c(a}D^{b)}\hat{\ep}_c\,,
\ee
together with ${\mathcal L}_{\hat{\ep}^c\partial_c}\sqrt{q}=(1/2)\sqrt{q}q^{ab}{\mathcal L}_{\hat{\ep}^c\partial_c}q_{ab}$, ${\mathcal L}_{\hat{\ep}^c\partial_c}q^{ab}=-q^{ac}q^{bd}{\mathcal L}_{\hat{\ep}^e\partial_e}q_{cd}$ and $\pi^{ab}=\sqrt{q}(K^{ab}-\lambda q^{ab}K)$, easily we reach the result
\bea
\label{IV-9}
{\mathcal L}_{\hat{\ep}^c\partial_c}\pi^{ab}&=&\pi^{ab}D_c\hat{\ep}^c+\sqrt{q}\left(\hat{\ep}^eD_eK^{ab}-2K^{e(a}D_e\hat{\ep}^{b)}
+2\lambda K D^{(a}\hat{\ep}^{b)}\right.\nonumber\\
&-&\left.2\lambda q^{ab} K^{cd}D_{(c}\hat{\ep}_{d)}-\lambda q^{ab}\hat{\ep}^eD_e K+2\lambda q^{ab}K^{cd}D_c\hat{\ep}^d\right)\nonumber\\
&=&\pi^{ab}D_c\hat{\ep}^c+\hat{\ep}^eD_e\pi^{ab}-2\pi^{e(a}D_e\hat{\ep}^{b)}\,.
\eea
Taking into account that the constraint in (\ref{IV-2b}) is the same vector constraint of GR, (\ref{IV-9}) reduces to~\cite{Thiemann}
\bea
\label{IV-10}
\delta_{\hat{\ep}^c\partial_c}\pi^{ab}&:=&\left\{\pi^{ab},H_c(\hat{\ep}^c)\right\}\nonumber\\
&=&{\mathcal L}_{\hat{\ep}^c\partial_c}\pi^{ab}\,.
\eea

So far, all the diffeomorphisms of GR are generated by the constraints of the Ho\v rava-Lifshitz theory. But things must change radically for the remaining case. It is already known from GR in the ADM formalism that the active diffeomorphism ${\mathcal L}_{Nn}$ may be generated by the constraint $H(N)$ only on shell. We shall show that it will also happen in the present case. The property
\be
\label{IV-11}
{\mathcal L}_{\hat{\ep}n}=\hat{\ep}{\mathcal L}_{n}=\ep^0{\mathcal L}_{Nn}
\ee
tells us that any diffeomorphism in the direction of $n$ may be generated by ${\mathcal L}_{Nn}$. By this reason, we may study the transformation $\delta_{Nn}\pi^{ab}$ only. The following straightforward result
\bea
\label{IV-12}
\delta_{Nn}\pi^{ab}&:=&\left\{\pi^{ab},H(N)\right\}\nonumber\\
&=&\frac{N}{2}q^{ab}{\mathcal H}-2N\sqrt{q}\left(K^{ac}K^b_c-\lambda K^{ab}K\right)\nonumber\\
&+&Nq^{ab}\left[R^{(3)}+f(R^{(3)},R^{(3)}_{ab},\dots)\right]+\sqrt{q}\left(D^aD^b-q^{ab}D^cD_c\right)N\nonumber\\
&+&\sqrt{q}\left[{\mathcal D}_f(N)\right]^{ab}-N\sqrt{q}R^{(3)ab}\,,
\eea
is obtained using the equality $\pi^{ac}\pi^b_c-\tilde{\lambda}\pi\pi^{ab}=q(K^{ac}K^b_c-\lambda K^{ab}K)$ after computing the Poisson bracket. In order to find ${\mathcal L}_{Nn}\pi^{ab}=N{\mathcal L}_{n}\pi^{ab}$, we must calculate ${\mathcal L}_n K_{\mu\nu}$ first~\footnote{Most of the details of these calculations can be found in \cite{Thiemann}.}, together with the Lie derivative property ${\mathcal L}_n K_{ab}={\mathcal L}_n (X^\mu_aX^\nu_bK_{\mu\nu})=X^\mu_aX^\nu_b{\mathcal L}_n K_{\mu\nu}$. It is interesting to begin with
\bea
\label{IV-13}
{\mathcal L}_n K_{\mu\nu}&=&n^\alpha\nabla_\alpha K_{\mu\nu}+2K_{\alpha(\mu}\nabla_{\nu)}n^{\alpha}\nonumber\\
&=&2K_{\alpha\mu}K^\alpha_\nu+q^\lambda_\mu q^\alpha_\nu \nabla_n K_{\lambda\alpha}\,.
\eea
From the relations that follow
\bea
\label{IV-14}
R^{(3)}_{\lambda\beta}&=&q^{\lambda^\prime}_\lambda q^{\beta^\prime}_\beta R^{(4)}_{\lambda^\prime\beta^\prime}
+q^{\lambda^\prime}_\lambda q^{\beta^\prime}_\beta n^{\sigma^\prime}[\nabla_{\lambda^\prime},\nabla_{\sigma^\prime}]n_{\beta^\prime}
-KK_{\lambda\beta}+K^\sigma_\lambda K_{\sigma\beta}\,,
\eea
\bea
\label{IV-15}
q^{\lambda^\prime}_\lambda q^{\beta^\prime}_\beta n^{\sigma^\prime}[\nabla_{\lambda^\prime},\nabla_{\sigma^\prime}]n_{\beta^\prime}
&=&q^{\lambda^\prime}_\lambda q^{\beta^\prime}_\beta\nabla_{\lambda^\prime}\nabla_n n_{\beta^\prime}
+\nabla_n n_\lambda\nabla_n n_\beta \nonumber\\
&-& q^{\lambda^\prime}_\lambda q^{\beta^\prime}_\beta\nabla_nK_{\lambda^\prime\beta^\prime}
-K^\sigma_\lambda K_{\beta\sigma}\,,
\eea
and also (see \cite{Thiemann} for more details)
\be
\label{IV-16}
\frac{1}{N}D_\lambda D_\beta N=q^{\lambda^\prime}_\lambda q^{\beta^\prime}_\beta\nabla_{\lambda^\prime}\nabla_n n_{\beta^\prime}
+\nabla_n n_\lambda\nabla_n n_\beta\,,
\ee
we obtain, already in the pullbach $\Re\times\sigma$, the desired result
\be
\label{IV-17}
{\mathcal L}_n K_{ab}=-KK_{ab}+2K_{ac}K^c_b-R^{(3)}_{ab}+q^c_aq^d_bR^{(4)}_{cd}+\frac{1}{N}D_aD_bN\,.
\ee
Omitting the details we quote
\bea
\label{IV-18}
{\mathcal L}_{Nn}\pi^{ab}&=&-2N\sqrt{q}\left(K^{ac}K_c^b-\lambda KK^{ab}\right)-N\sqrt{q}\left(R^{(3)ab}-\lambda q^{ab}R^{(3)}\right)\nonumber\\
&+&\sqrt{q}\left(D^aD^b-q^{ab}D^cD_c\right)N+N\sqrt{q}\left(q^{ac}q^{bd}-\lambda q^{ab}q^{cd}\right)R^{(4)}_{cd}\,.
\eea

At this point we can join the result in (\ref{IV-12}) with the one in (\ref{IV-18}) to write
\bea
\label{IV-19}
\delta_{Nn}\pi^{ab}&=&\frac{N}{2}q^{ab}{\mathcal H}+Nq^{ab}\left[(1-\lambda)R^{(3)}+f(R^{(3)},R^{(3)}_{ab},\dots)\right]\nonumber\\
&-&\sqrt{q}q^{ab}(1-\lambda)D^cD_cN+\sqrt{q}\left[{\mathcal D}_f(N)\right]^{ab}\nonumber\\
&-&N\sqrt{q}\left(q^{ac}q^{bd}-\lambda q^{ab}q^{cd}\right)R^{(4)}_{cd}+{\mathcal L}_{Nn}\pi^{ab}\,.
\eea
Clearly, $H(N)$ does not generate any diffeomorphism on $\pi^{ab}$, at least off shell, as well as it happens in GR. However, in the case of (\ref{IV-19}), we hardly expect that it will occur on shell. The presence of the function  $f$ and the parameter $\lambda$ appears to make it just impossible. We shall prove that, in fact, the transformation $\delta_{Nn}\pi^{ab}$ reduces to ${\mathcal L}_n\pi^{ab}$ on shell.

From now on we will be working on shell ($\tilde{G}_{ab}=0$.) We call the relation $g^{\mu\nu}G_{\mu\nu}=-R^{(4)}$ that, together with (\ref{21}), result into
\bea
\label{IV-20}
R^{(4)}_{ab}&=&\frac{q_{ab}}{2}\bigg[(1-\lambda)K^2-f(R^{(3)},R^{(3)}_{ab},\dots)+(1-\lambda)\nabla_n K\nonumber\\
&&\left.-\frac{1}{N}q^{cd}\left[{\mathcal D}_f(N)\right]_{cd}\right]+\frac{1}{N}\left[{\mathcal D}_f(N)\right]_{ab}\,.
\eea
Collecting (\ref{IV-13}), (\ref{IV-17}), and also the property $n^\mu n^\nu\nabla_n K_{\mu\nu}=0\Rightarrow q^{\mu\nu}\nabla_n K_{\mu\nu}=\nabla_nK$, one obtains
\be
\label{IV-21}
\nabla_nK=-R^{(3)}+q^{ab}R^{(4)}_{ab}+\frac{1}{N}D^aD_a N-K^2\,.
\ee
From (\ref{IV-20}), we may rewrite (\ref{IV-21}) as
\bea
\label{IV-22}
(3\lambda-1)\nabla_nK&=&-2R^{(3)}-(3\lambda-1)K^2-3f(R^{(3)},R^{(3)}_{ab},\dots)\nonumber\\
&-&\frac{1}{N}q^{ab}\left[{\mathcal D}_f(N)\right]_{ab}
+\frac{2}{N}D^aD_aN\,.
\eea
After a series of cancellations we get
\bea
\label{IV-23}
\fl(q^{ac}q^{bd}-\lambda q^{ab}q^{cd})R^{(4)}_{cd}=\frac{1}{N}\left[{\mathcal D}_f(N)\right]^{ab}\nonumber\\
+q^{ab}\left[(1-\lambda)R^{(3)}+f(R^{(3)},R^{(3)}_{ab},\dots)-\frac{(1-\lambda)}{N}D^cD_cN\right]\,,
\eea
what lead us to
\be
\label{IV-24}
\left.\delta_{Nn}\pi^{ab}\right|_{OS}={\mathcal L}_{Nn}\pi^{ab}\,,
\ee
on shell. Since the action, not the Lagrangian, is invariant under active diffeomorphism, the Hamiltonian and vector constraint are the generators of all the diffeomorphisms of the group Diff(M).

The result obtained above is really interesting. It tells us that among the various forms of modification of general relativity that are diffeomorphism invariant, there exists a subset of modified theories, characterized here by the term containing $\lambda$ and the function f, that generate all the diffeomorphisms of GR, as in the ADM formalism. Of course it does not mean that these modified theories are consistent in other respects. We are not sure even if the number of degrees of freedom remain two. We also did not verify if the constraints of the theory remain first class. In the case where the constraints become second class, we may obtain some active diffeomorphism transformations that are not symmetry transformations as well as is the case of Einstein theory. The implications of this property could be in the heart of several ills~\cite{Padilla1,Padilla2,Padilla3} that arise in the Horava-Lifshitz gravity.

\section{Conclusions}
\label{Conclusions}

We have demonstrated that the Ho\v rava-Lifshitz gravity does not lose diffeomorphism invariance due to the presence of the parameter $\lambda$. However, we have showed that the terms containing the constant Levi-Civita symbol, present in the original proposal of Ho\v rava, turn the theory dependent of any diffeomorphism. A modified theory of gravity should at least preserve a certain class of diffeomorphisms. Then, we neglected those terms and showed that the theory is invariant under the whole group of diffeomorphisms of General Relativity. Next, we have calculated the equations of motion and displayed their direct relation with the Hamiltonian and vector constraints. Finally, we have proved that the set of modifications of gravity labeled by $\lambda$ and the class of functions $f$, in the Hamiltonian formalism, generates all the diffeomorphisms of General Relativity on the energy shell, as well as it occurs in the ADM original formalism.

\ack
This work was partially supported by Fundação de Amparo à Pesquisa do Estado de São Paulo (FAPESP).


\section*{References}

\begin{thebibliography}{10}

\bibitem{Horava2}
Horava P 2009 {\it Phys. Rev.} D {\bf 79} 084008

\bibitem{Sotiriou}
Sotiriou T P 2010 (arXiv:hep-th/1010.3218)

\bibitem{Padilla1}
Charmousis C, Niz G, Padilla A, and Saffin P M 2009 {\it J. High Energy Phys.} JHEP08(2009)070

\bibitem{Padilla2}
Kimpton I and Padilla A 2010 {\it J. High Energy Phys.} JHEP07(2010)014

\bibitem{Padilla3}
Padilla A 2010 (arXiv:hep-th/1009.4074)

\bibitem{Germani}
Germani C, Kehagias A, and Sfetsos K 2009 {\it J. High Energy Phys.} JHEP09(2009)060

\bibitem{Horava1}
Horava P 2009 {\it J. High Energy Phys.} JHEP03(2009)020

\bibitem{Bogdanos}
Bogdanos C and Saridakis E N 2010 {\it Class. Quant. Grav.} {\bf 27} 075005

\bibitem{Henneaux}
Henneaux M, Kleinschmidt A, and Gomez GL 2010 {\it Phys. Rev.} D {\bf 81} 064002

\bibitem{Pons}
Pons J M and Talavera P 2010 {\it Phys. Rev.} D {\bf 82} 044011

\bibitem{Bellorin}
Bellorin J and Restuccia A 2010 (arXiv:hep-th/1004.0055)

\bibitem{Stelle}
Stelle K S 1978 {\it Gen. Rel. Grav.} {\bf 9} 353--371

\bibitem{Thiemann}
Thiemann T 2007 {\it Modern Canonical Quantum General Relativity} 1st edn (Cambridge: Cambridge University Press)

\bibitem{Wald}
Wald R M 1984 {\it General Ralativity}.
(Chicago: The University of Chicaco Press)

\bibitem{Senhoras}
Choquet-Bruhat Y, DeWitt-Morette C, and Dillard-Bleick M 1996 {\it Analysis, Manifolds and Physics} (Amsterdam: North-Holland/American
Elsevier)

\bibitem{Dirac1}
Dirac P A M 2001 {\it Lectures on Quantum Mechanics} (Dover Publications)

\end{thebibliography}

\end{document}